\documentclass{article}

\input{tcilatex}

\begin{document}

\pagestyle{myheadings} \markright{\it 241-1} \vskip.5in

\begin{center}
%
%
\vskip.4in \textbf{Light-front time picture of the Bethe-Salpeter equation
fermionic}{\Large \textbf{\ }} \vskip.3in 
%
%
%
J.H.O.Sales\footnote{%
Email: \texttt{henrique@fatecsp.br }}\\[0pt]
Faculdade de Tecnologia de S\~{a}o Paulo-DEG, P\c{c}a. Coronel Fernando
Prestes, 01124-060 S\~{a}o Paulo-SP. %
%
\end{center}

%
The reduction of the equation of Bethe-Salpeter of two fermions in front of
light is studied for the Yukawa model. We use the light-front Green's
function for the N-particle system for two-fermions plus N-2 intermediate
bosons.\vskip.2in

%
%
%
%
%
%
%
%
%
%
%

\section{Light-front}

The light-front coordinates are defined in terms of these by the following
relations: 
\begin{equation}
x^{+}=x^{0}+x^{3},\text{ }x^{-}=x^{0}-x^{3}\text{ and }\vec{x}^{\perp }=x^{1}%
\vec{i}+x^{2}\vec{j},
\end{equation}
where $\vec{i}$ and $\vec{j}$ are the unit vectors in the direction of the
coordinates $x$ and $y$. The null plane is defined by $x^{+}=0$, that is,
this condition defines the hyper-surface which is tangent to the light cone,
the reason why some authors call those light-cone coordinates.

Note that for the usual four-dimensional Minkowski space-time whose metric $%
g^{\mu\nu}$ is defined such that its signature is $(1,-1,-1,-1)$ we have 
\begin{eqnarray}
x^{+}&=&x^{0}+x^{3}\;\;=\;\;x_0-x_3\;\;\equiv \;\;x_{-},  \nonumber \\
x^{-}&=&x^{0}-x^{3}\;\;=\;\;x_0+x_3\;\;\equiv \;\;x_{+},  \nonumber \\
\vec{x}^{\perp }&=& x^{1}\vec{i}+x^{2}\vec{j}\;\;=\;\;-x_{1}\vec{i}-x_2\vec{j%
}\;\;\equiv \;\;-x_{\perp},
\end{eqnarray}

The initial boundary conditions for the dynamics in the light front are
defined in this hyper-plane. Note that the axis $x^{+}$ is orthogonal to the
plane $x^{+}=0$. Therefore, a displacement of this hyper-surface for $%
x^{+}>0 $ is analogous to the displacement of the plane $t=0$ for $t>0$ of
the usual four-dimensional space-time. With this analogy we identify $x^{+}$
as the ``time'' coordinate for the null plane. Of course, since there is a
conspicuous discrete symmetry between $x^+ \leftrightarrow x^-$, one could
choose $x^-$ as his ``time'' coordinate. However, once chosen, one has to
stick to the convention adopted. We shall adhere to the former one.

The canonically conjugate momenta for the coordinates $x^{+},x^{-}$ and $%
x^{\perp }$ are defined respectively by: 
\begin{equation}
k^{+}=k^{0}+k^{3},\text{ }k^{-}=k^{0}-k^{3}\text{ and }k^{\perp }=\left(
k^{1},k^{2}\right) .  \label{1.4}
\end{equation}

The scalar product in the light front coordinates becomes therefore 
\begin{equation}
a^{\mu }b_{\mu }=\frac{1}{2}\left( a^{+}b^{-}+a^{-}b^{+}\right) - \vec{a}%
^{\perp }\cdot \vec{b}^{\perp },  \label{1.7}
\end{equation}
where $\vec{a}^{\perp }$ and $\vec{b}^{\perp }$ are the transverse
components of the four vectors. All four vectors, tensors and other entities
bearing space-time indices such as Dirac matrices $\gamma ^{\mu }$ can be
expressed in this new way, using components $(+,-,\perp)$.

From (\ref{1.7}) we can get the scalar product $x^{\mu }k_{\mu }$ in the
light front coordinates as $x^{\mu }k_{\mu }=\frac{1}{2}\left(
x^{+}k^{-}+x^{-}k^{+}\right) -\vec{x}^{\perp }\cdot \vec{k}^{\perp }.$

Here again, in analogy to the usual four-dimensional Minkowski space-time
where such a scalar product is $x^{\mu }k_{\mu }=x^{0}k^{0}-\mathbf{{x}\cdot 
{k}}$ where $\mathbf{x}$ is the three-dimensional vector, with the energy $%
k^{0}$ associated to the time coordinate $x^{0}$, we have the light-front
``energy'' $k^{-}$ associated to the light-front ``time'' $x^{+}$. Note,
however, that there is a crucial difference between the two formulations:
while the usual four-dimensional space-time is Minkowskian, the light-front
coordinates projects this onto two sectorized Euclidean spaces, namely $%
(+,-) $, and $(\perp ,\perp )$.

In the Minkowski space described by the usual space-time coordinates we have
the relation between the rest mass and the energy for the free particle
given by $k^{\mu }k_{\mu }=m^{2}$. Using (\ref{1.7}), we have $k^{\mu
}k_{\mu }=\frac{1}{2}\left( k^{+}k^{-}+k^{-}k^{+}\right) -\vec{k}^{\perp
}\cdot \vec{k}^{\perp },$ so that 
\begin{equation}
k^{-}=\frac{\overrightarrow{k}_{\perp }^{2}+m^{2}}{k^{+}}.  \label{1.10}
\end{equation}

Note that the energy of a free particle is given by $k^{0}=\pm \sqrt{m^{2}+%
\mathbf{{k}^{2}}}$, which shows us a quadratic dependence of $k^{0}$ with
respect to $\mathbf{k}$. These positive/negative energy possibilities for
such a relation were the source of much difficulty in the interpretation of
the negative energy particle states in the beginning of the quantum field
theory descriprion for particles, finally solved by the antiparticle
interpretation given by Feynman. In contrast to this, we have a linear
dependence between $(k^{+})^{-1}$ and $k^{-}$ (see Eq.(\ref{1.10})), which
immediately reminds us of the non-relativistic quantum mechanical type of
relationship for one particle state systems.

\section{Boson and fermion propagator}

The 1-body Green's functions can be derived from the covariant propagator
for 1-particles propagating at equal light-front times. In this case the
propagator from $x^{+}=0$ to $x^{+}>0$ is given by

\begin{equation}
S(x^{+})=\frac{1}{2}\int \frac{dk^{-}dk^{+}dk^{\perp }}{\left( 2\pi \right) }%
\frac{ie^{\frac{-i}{2}k^{-}x^{+}}}{k^{+}\left( k^{-}-\frac{k_{\perp
}^{2}+m^{2}}{k^{+}}+\frac{i\varepsilon }{k^{+}}\right) }.  \label{2b}
\end{equation}

The Fourier transform to the total light-front energy $(P^{-})$ is given by $%
{S}(P^{-})=\frac{1}{2}\int dx^{+}e^{\frac{i}{2}P^{-}x^{+}}S(x^{+})$ and the
free 1-body Green's function is given by $S(k^{-})=\frac{1}{k^{+}}G(k^{-})$,
where$\ $%
\begin{equation}
G_{0}^{(1)}(k^{-})=\frac{\theta (k^{+})}{k^{-}-k_{on}^{-}}\   \label{4b}
\end{equation}
with $k_{on}^{-}=\frac{k_{\perp }^{2}+m^{2}}{k^{+}}$ being the light-front
Hamiltonian of the free 1-particle system.

Let $S_{\text{F}}$ denote fermion field propagator in covariant theory 
\begin{equation}
S_{\text{F}}(x^{\mu })=\int \frac{d^{4}k}{\left( 2\pi \right) ^{4}}\frac{i(%
\rlap\slash k_{\text{on}}+m)}{k^{2}-m^{2}+i\varepsilon }e^{-ik^{\mu }x_{\mu
}},  \label{4}
\end{equation}
where $\rlap\slash k_{\text{on}}=\frac{1}{2}\gamma ^{+}\frac{(k^{\perp
})^{2}+m^{2}}{k^{+2}}+\frac{1}{2}\gamma ^{-}k^{+}-\gamma ^{\perp }k^{\perp }$%
. Using light-front variables in the Eq.(\ref{4}), we have 
\begin{equation}
S_{\text{F}}(x^{+})=\frac{i}{2}\int \frac{dk^{-}dk^{+}dk^{\perp }}{\left(
2\pi \right) }\left[ \frac{\rlap\slash k_{on}+m}{k^{+}\left(
k^{-}-k_{on}^{-}+\frac{i\varepsilon }{k^{+}}\right) }+\frac{\gamma ^{+}}{%
2k^{+}}\right] e^{\frac{-i}{2}k^{-}x^{+}}.  \label{5}
\end{equation}

We note that for the fermion field, light-front propagator differs from the
Feynmam propagator by an instantaneous propagator.

The free 1-fermion Green's function is given by 
\begin{equation}
G(k^{-})=\frac{\Lambda _{+}\left( k\right) }{\left( k^{-}-k_{on}^{-}+\frac{%
i\varepsilon }{k^{+}}\right) },  \label{7.21}
\end{equation}
where $\Lambda _{\pm }\left( k\right) =\frac{\pm \rlap\slash k_{on}+m}{2m}%
\theta (\pm k^{+})$.

\section{Coupled equations for the Green's functions}

The light-front Green's function for the two fermions system obtained from
the solution of the covariant BS equation that contains all two-body
irreducible diagrams, with the exception of those including closed loops of
bosons $\Psi _{1}$ and $\Psi _{2}$ and part of the cross-ladder diagrams, is
given by: 
\begin{equation}
G^{(2)}(K^{-})=G_{0}^{(2)}(K^{-})+G_{0}^{(2)}(K^{-})VG^{(3)}(K^{-})VG^{(2)}(K^{-})\ ,
\label{3.28}
\end{equation}
\begin{equation}
G^{(3)}(K^{-})=G_{0}^{(3)}(K^{-})+G_{0}^{(3)}(K^{-})VG_{0}^{(4)}(K^{-})VG^{(3)}(K^{-}).
\label{3.29}
\end{equation}
In the Yukawa model for fermions, the interaction operator acting between
Fock-states differing by zero, one and two $\sigma $'s, has matrix elements
given by 
\begin{eqnarray}
&&\langle (q,s^{\prime })k_{\sigma }|V|(k,s)\rangle =-2m(2\pi )^{3}\delta
(q+k_{\sigma }-k)\frac{g_{S}}{\sqrt{q^{+}k_{\sigma }^{+}k^{+}}}\theta
(k_{\sigma }^{+}){\overline{u}}(q,s^{\prime })u(k,s)~,  \label{mia} \\
&&\langle (q,s^{\prime })k_{\sigma }^{\prime }|V|(k,s)k_{\sigma }\rangle
=-2(2\pi )^{3}\delta (q+k_{\sigma }^{\prime }-k-k_{\sigma })\delta
_{s^{\prime }s}\frac{g_{S}^{2}}{\sqrt{k_{\sigma }^{\prime +}k_{\sigma }^{+}}}%
{\frac{\theta (k_{\sigma }^{\prime +})\theta (k_{\sigma }^{+})}{%
k^{+}+k_{\sigma }^{+}}}~,  \label{mib} \\
&&\langle (q,s^{\prime })k_{\sigma }^{\prime }k_{\sigma }|V|(k,s)\rangle
=-2(2\pi )^{3}\delta (q+k_{\sigma }^{\prime }+k_{\sigma }-k)\delta
_{s^{\prime }s}\frac{g_{S}^{2}}{\sqrt{k_{\sigma }^{\prime +}k_{\sigma }^{+}}}%
{\frac{\theta (k_{\sigma }^{\prime +})\theta (k_{\sigma }^{+})}{%
k^{+}-k_{\sigma }^{+}}}~.  \label{mic}
\end{eqnarray}
The instantaneous terms in the two-fermion propagator give origin to Eqs. (%
\ref{mib}) and (\ref{mic}).

A systematic expansion by the consistent truncation of the light-front Fock
space up to $N$ particles in the intermediate states (boson 1, boson 2 and $%
N-2$ $\sigma $'s) in the set of Eqs.(\ref{3.29}) and (\ref{3.28}), amounts
to substitution $G^{(3)}(K^{-})\cong G_{0}^{(3)}(K^{-})$. The kernel of Eq.(%
\ref{3.28}) still contains an infinite sum of light-front diagrams, that are
obtained solving by Eq.(\ref{3.29}). To obtain the ladder aproximation up to
order $g^{2}$, Eq.(\ref{3.28}), only the free and first order terms are kept
in Eq.(\ref{3.29}), with the restriction of only one and one boson covariant
exchanges.Therefore, we have for Eq.(\ref{3.28}) 
\begin{equation}
G_{g^{2}}^{(2)}(K^{-})=G_{0}^{(2)}(K^{-})+G_{0}^{(2)}(K^{-})VG_{0}^{(3)}(K^{-})VG_{g^{2}}^{(2)}(K^{-}),
\label{5.1a}
\end{equation}
Taking the two-boson system as an example and restricting the intermediate
state propagation up to 3-particles, we find that 
\begin{equation}
G_{g^{2}}^{(2)}(K^{-})=G_{0}^{(2)}(K^{-})+G_{0}^{(2)}(K^{-})VG_{0}^{(3)}(K^{-})V\left\{ G_{0}^{(2)}(K^{-})+G_{0}^{(2)}(K^{-})VG_{0}^{(3)}(K^{-})VG_{g^{2}}^{(2)}(K^{-})\right\}
\label{3.29a}
\end{equation}
The correction in order $g^{2}$ is given for $\Delta
G_{g^{2}}^{(2)}(K^{-})=G_{0}^{(2)}(K^{-})VG_{0}^{(3)}(K^{-})VG_{0}^{(2)}(K^{-}) 
$.

\section{Bethe-Salpeter equation}

We perform the quasi-potential reduction of two-body BSE's and present the
coupled set of equations for the light-front Green's functions for bosonic
and fermionic models, with the interaction Lagrangian respectively given by, 
$\mathcal{L}_{I}^{B}=g_{S}\phi _{1}^{\dagger }\phi _{1}\sigma +g_{S}\phi
_{2}^{\dagger }\phi _{2}\sigma $ and $\mathcal{L}_{I}^{F}=g_{S}\overline{%
\Psi }\Psi \sigma $, where $\phi _{1}$, $\phi _{2}$ and $\sigma $ are the
bosonic fields and $\Psi $ is the fermion field in the Yukawa model.

Close the area of energy of the bound state the Green's function has a pole $%
\lim_{K^{-}\rightarrow K_{B}^{-}}G^{\left( 2\right) }(K^{-})=\frac{\left|
\psi _{B}\right\rangle \left\langle \psi _{B}\right| }{K^{-}-K_{B}^{-}}$,
where $\left| \psi _{B}\right\rangle $ it is the wave-function of the bound
state.

The homogeneous equation for the light-front two-body bound state
wave-function is obtained the solution of 
\begin{equation}
|\Psi _{B}>=G_{0}^{(2)}(K_{B}^{-})VG^{(3)}(K_{B}^{-})V|\Psi _{B}>,
\label{5.3}
\end{equation}

The vertex function for the bound state wave-function is defined as 
\begin{equation}
\Gamma _{B}(k_{\perp },q^{+})=<k,\ K-k|\left( G_{0}^{(2)}(K_{B}^{-})\right)
^{-1}|\Psi _{B}>.  \label{vertexlf}
\end{equation}

The Green's function obtained from this equation, up to order $g^{2}$,
reproduces the covariant two-body propagator between two light-front
hypersurfaces. In this approximation, the vertex function satisfies the
following integral equation, 
\begin{equation}
\Gamma _{B}(\overrightarrow{q}_{\perp },y)=\int \frac{dxd^{2}k_{\perp }}{%
x(1-x)}\frac{\mathit{K}^{(3)}(\overrightarrow{q}_{\perp },y;\overrightarrow{k%
}_{\perp },x)}{M_{B}^{2}-M_{0}^{2}}\Gamma _{B}(\overrightarrow{k}_{\perp
},x),  \label{5.7}
\end{equation}
where the momentum fractions are $y=q^{+}/K^{+}$ and $x=k^{+}/K^{+}$, with $%
0<y<1$. Where $\overrightarrow{K}_{\perp }=0$, and $%
M_{0}^{2}=K^{+}K_{(2)on}^{-}-K_{\perp }^{2}=\frac{k_{\perp }^{2}+m^{2}}{%
x(1-x)}$. The part of the kernel which contains only the propagation of
virtual three particle states foward in the light-front time is obtained
from Eq.(\ref{5.7}) as, 
\begin{eqnarray}
\mathcal{K}^{(3)}(y,q_{\perp };x,k_{\perp }) &=&\frac{g^{2}}{16\pi ^{3}}%
\frac{\Lambda _{+(1)}(q)\Lambda _{+(1)}(k)\Lambda _{+(2)}(K-q)\Lambda
_{+(2)}(K-k)\theta (y-x)}{\left( x-y\right) \left( M_{B}^{2}-\frac{%
\overrightarrow{q}_{\perp }^{2}+m^{2}}{1-y}-\frac{\vec{k}_{\perp }^{2}+m^{2}%
}{x}-\frac{(\overrightarrow{q}_{\perp }-\vec{k}_{\perp })^{2}+\mu ^{2}}{y-x}%
\right) }+  \nonumber \\
&&+\left[ k\leftrightarrow q\right] ,  \nonumber
\end{eqnarray}
being $M_{B}^{2}=K_{B}^{+}K_{B}^{-}$. W got the attention for the
relationship between $\left| \Gamma _{B}\right\rangle $ and $\Gamma _{B}(%
\overrightarrow{q}_{\perp },y)$, defined for $\Gamma _{B}(\overrightarrow{q}%
_{\perp },y)=\sqrt{q^{+}(K^{+}-q^{+})}\left\langle \overrightarrow{q}_{\perp
},q^{+}\right. \left| \Gamma _{B}\right\rangle $.

The generalization for any order $g^{n}$ was calculated by the author \cite
{3}.

\section{Conclusion}

We formulated the Bethe-Salpeter equation in the ladder aproximation on the
light-front. The ladder comes from the exchange of an intermediate boson
between the other two fermions. The kernel of the integral equation is
constructed from the perturbative light-front propagator of the interacting
two fermions system up to $\mathcal{O}(g^{2}).$

We can use that method for the study of coupled virtual gauge boson \cite{4}.

\textbf{\textit{Acknowledgments}:} J.H.O.Sales thank support from FAPESP
(S\~{a}o Paulo, Brasil)

\end{document}